\documentclass[aps,prl,reprint,superscriptaddress,nofootinbib,longbibliography,floatfix]{revtex4-2}

\usepackage{amsmath,amssymb,mathtools,bm}
\usepackage{graphicx}
\usepackage{xcolor}
\definecolor{ZZGNote}{RGB}{0,105,85}
\usepackage[colorlinks=true,urlcolor=blue,linkcolor=blue,citecolor=blue]{hyperref}

\graphicspath{{figures/}{./}}

\newcommand{\kk}{\mathbf{k}}

\newcommand{\qq}{\mathbf{q}}
\newcommand{\RR}{\mathbf{R}}
\newcommand{\Tr}{\mathrm{Tr}}
\newcommand{\Ree}{\mathrm{Re}}
\newcommand{\cP}{\mathcal{P}}
\newcommand{\cR}{\mathcal{R}}

\newcommand{\xiR}{\xi_{\mathrm{RKKY}}}

\begin{document}

\title{Quantum Geometry-Driven RKKY: From Flat to Dispersive Bands}

\author{Zhenggang Zhou}
\affiliation{Center for Correlated Matter and School of Physics, Zhejiang University, Hangzhou 310058, China}

\author{Chuang Li}
\affiliation{Center for Correlated Matter and School of Physics, Zhejiang University, Hangzhou 310058, China}

\author{Lun-Hui Hu}
\email{lunhui@zju.edu.cn}
\affiliation{Center for Correlated Matter and School of Physics, Zhejiang University, Hangzhou 310058, China}

\begin{abstract}
In flat-band systems, quantum metric bounds physical observables like superfluid weight and coherence length, suggesting a single geometric scale for spatial correlations. Here, we show that the RKKY exchange in an isolated filled flat band can violate this expectation. With the intraband channel absent, the exchange proceeds via virtual interband transitions across the gap; the kernel becomes the inverse-gap-weighted trace product of the real-space flat-band projector and empty-band projectors. Because the corresponding momentum-space projectors are analytic, the kernel decays exponentially, with a decay length $\xi_\text{RKKY}$ set by the closest singularities of the analytically continued projectors in the complex momentum plane. Thus, this length depends on both the flat-band geometry and the band gap. Applying this formalism to Chern flat-band systems, we find that for small gaps, $\xi_\text{RKKY}$ depends non-monotonically on the quantum metric length: it first decreases, then increases, revealing that stronger quantum geometry can shorten the magnetic exchange range. Upon restoring dispersion, a nontrivial inversion representation can force the overlap between Bloch states at antipodal Fermi points to vanish under gate tuning, producing a $1/R^3$ RKKY tail instead of the conventional $1/R^2$---a geometric selection effect.
\end{abstract}

\maketitle

\textit{Introduction.---}
Flat bands provide a fertile playground for studying ferromagnetism under the Hubbard interaction~\cite{mielke1993ferromagnetism}. Their recent identification in Moir\'e systems~\cite{bistritzer2011moire,cao2018unconventional,wu2019prl,tang2020simulation} and thousands of stoichiometric crystalline materials~\cite{regnault2022catalogue} offers a platform for exploring quantum geometric effects in real materials. With the kinetic energy quenched, even weak interactions can dominate, giving rise to correlated quantum phases such as unconventional superconductivity~\cite{kopnin2011prb,peotta2015superfluidity,Julku2016prl,Liang2017prb}, fractional Chern insulators~\cite{Regnault2011prx,Neupert2011prl,Sun2011prl,tang2011prl,sheng2011fractional}, and correlated insulating states~\cite{wu2007prl}. In such systems, the quantum geometry—encompassing quantum metric and Berry curvature—governs both correlated phases and transport~\cite{paivi2023prl,yu2025quantum,gao2025quantum,liu2025quantum,jiang2025revealing,verma2026quantum,kitamura2026quantum}. This is particularly evident in flat-band superconductors, where the superfluid weight is determined by the quantum metric rather than the conventional density-over-mass ratio~\cite{Kukka2022prb,tian2023evidence,tanaka2025superfluid}, and the coherence length and vortex size are bounded by the quantum metric length scale~\cite{Chen2024prl,Iskin2023prb,hu2025anomalous,li2025vortex}. These findings suggest that a single geometric scale, the quantum metric length~\cite{Marzari1997prb}, may universally control spatial correlations in flat-band systems. However, whether this picture holds for all correlation functions, particularly those involving localized magnetic moments, remains an open question.

This question is naturally addressed by the RKKY exchange, which probes real-space spin correlations mediated by itinerant electrons~\cite{coleman2015introduction}. For a spin-rotation-invariant Kondo coupling, the band-resolved static kernel for the RKKY exchange reads
\begin{align} \label{eq:intro_kernel}
X(\qq)= \sum_{mn}\int_{\kk} d^2\kk \, \frac{f_{m,\kk} - f_{n,\kk+\qq}}{E_{n,\kk+\qq} -E_{m,\kk}} F_{mn}(\kk,\qq),
\end{align}
where $E_{m,\kk}$ are band energies, $f_{m,\kk}$ are Fermi-Dirac distribution functions, and $F_{mn}(\kk,\qq)=\Tr[P_m(\kk)P_n(\kk+\qq)]$ is the overlap of band projectors $P_m(\kk)$ and $P_n(\kk+\qq)$. The real-space RKKY exchange follows from $X(\RR)=\int_{\qq} d^2\qq \, e^{i\qq\cdot\RR}X(\qq)$. This kernel cleanly separates the spectral particle--hole propagation from the projector overlap. For a simple parabolic-band metal, where the projector overlap is constant, this reduces to the Lindhard response, whose $2k_F$ nonanalyticity fixes the conventional $1/R^2$ asymptotic tail in two dimensions. How does this picture change when the band is flat and fully filled, so that the conventional $2k_F$ mechanism is absent and the response is instead controlled by the band geometry?

Here, we show that the RKKY exchange in a filled flat band decays exponentially, $X(\RR) \propto e^{-R/\xi_{\rm RKKY}}$, and that $\xi_{\rm RKKY}$ can violate the single quantum metric length-scale paradigm. This length is set not by the quantum metric alone, but by the closest singularities of the band projectors upon analytic continuation to complex momentum, leading to a non-monotonic dependence on the quantum metric weight. Counterintuitively, for small gaps, increasing the quantum geometric weight can shorten the exchange range. We demonstrate this analytically for ideal Chern flat bands, where the violation is particularly pronounced when the gap to the dispersive bands is small. Upon restoring dispersion, the exponential tail crosses over to algebraic decay; the nontrivial inversion representation then enforces a $1/R^3$ RKKY tail instead of the conventional $1/R^2$, with gate tuning providing experimental control over this geometric selection effect.

\textit{General analysis.---}
We now specialize Eq.~\eqref{eq:intro_kernel} to an isolated flat band that is fully occupied. Take $E_{m,\kk} \equiv E_o \leq E_F$ denote the flat-band energy below the Fermi energy. Due to the band flatness, the intra-band channel vanishes, implying that the flat-band RKKY is driven solely by inter-band contributions that encode the quantum geometry. The spectral factor depends only on $\kk+\qq$, thus $X(\RR)$ becomes the geometric RKKY kernel,
\begin{align} \label{eq:projector_resolvent}
X(\RR) = 2\,\Ree\, \Tr\! \left[ \cP_o(\RR) \, \cR_u(\RR)
\right],
\end{align}
where $\cP_o(\RR)=\int_{\kk} d^2\kk\, e^{-i\kk\cdot\RR}P_o(\kk)$ is the real-space projector onto the flat-band subspace, and $\cR_u(\RR) = \sum_u\int_{\kk} d^2\kk \, e^{i\kk\cdot\RR}P_u(\kk)/(E_u(\kk)-E_o)$ with $E_u(\kk)\geq E_F$ is the inverse-gap-weighted real-space projector onto the remote unoccupied bands. The trace is over internal degrees of freedom.

We first note that both $\cP_o(\RR)$ and $\cR_u(\RR)$ decay exponentially as $R\to\infty$. For a Chern flat band, $\cP_o(\RR)$ is structurally analogous to the Landau level projector, which decays exponentially with a length set by the magnetic length~\cite{gao2025quantum}. More generally, any isolated flat-band projector decays exponentially in real space, since it is the Fourier transform of an analytically continued band projector; analyticity follows from Kato's perturbation theory for any finite spectral gap~\cite{kato1966perturbation}, regardless of the Chern number. The unoccupied resolvent $\cR_u(\RR)$ decays with a similar exponential envelope. Unlike in the free-electron case, the energy denominators $E_u(\kk)-E_o$ possess no real zeros due to the finite gap. At large $R$, the contour is deformed into the complex plane, where the momentum acquires an imaginary part and $e^{i\kk\cdot\RR}$ becomes exponentially suppressed. Thus, the flat-band RKKY exchange is driven by quantum geometry and decays exponentially in real space—a hallmark of the gapped interband particle-hole continuum, where the absence of low-energy excitations replaces the conventional $1/R^2$ metallic tail with a short-ranged exchange.

In general, the geometric RKKY kernel in Eq.~\eqref{eq:projector_resolvent} takes the asymptotic form $X(\RR) \propto \exp(-R/\xi_{\rm RKKY})$, which defines the decay length $\xi_{\rm RKKY}$. Since the kernel is the trace of a matrix product of $\cP_o(\RR)$ and $\cR_u(\RR)$, the inverse decay length naturally decomposes as
\begin{align} \label{eq:xi-RKKY}
\xi_{\rm RKKY}^{-1} = \xi_o^{-1} + \xi_u^{-1},
\end{align}
where $\cP_o(\RR) \simeq \exp(-R/\xi_o)$ and $\cR_u(\RR) \simeq \exp(-R/\xi_u)$. Here, $\xi_o$ may be a quantum-geometric length scale inherited from the analytic structure of the flat-band projector, while $\xi_u$ is controlled primarily by the band gap. Consequently, $\xi_{\rm RKKY}$ is not merely the quantum metric length; it is a distinct geometric scale that also encodes the gap and the unoccupied-band contributions. It is instructive to see how both $\xi_o$ and $\xi_u$ contribute to $\xi_{\rm RKKY}$, a question we now address in a concrete model.

\begin{figure}[t]
\centering
\includegraphics[width=\columnwidth]{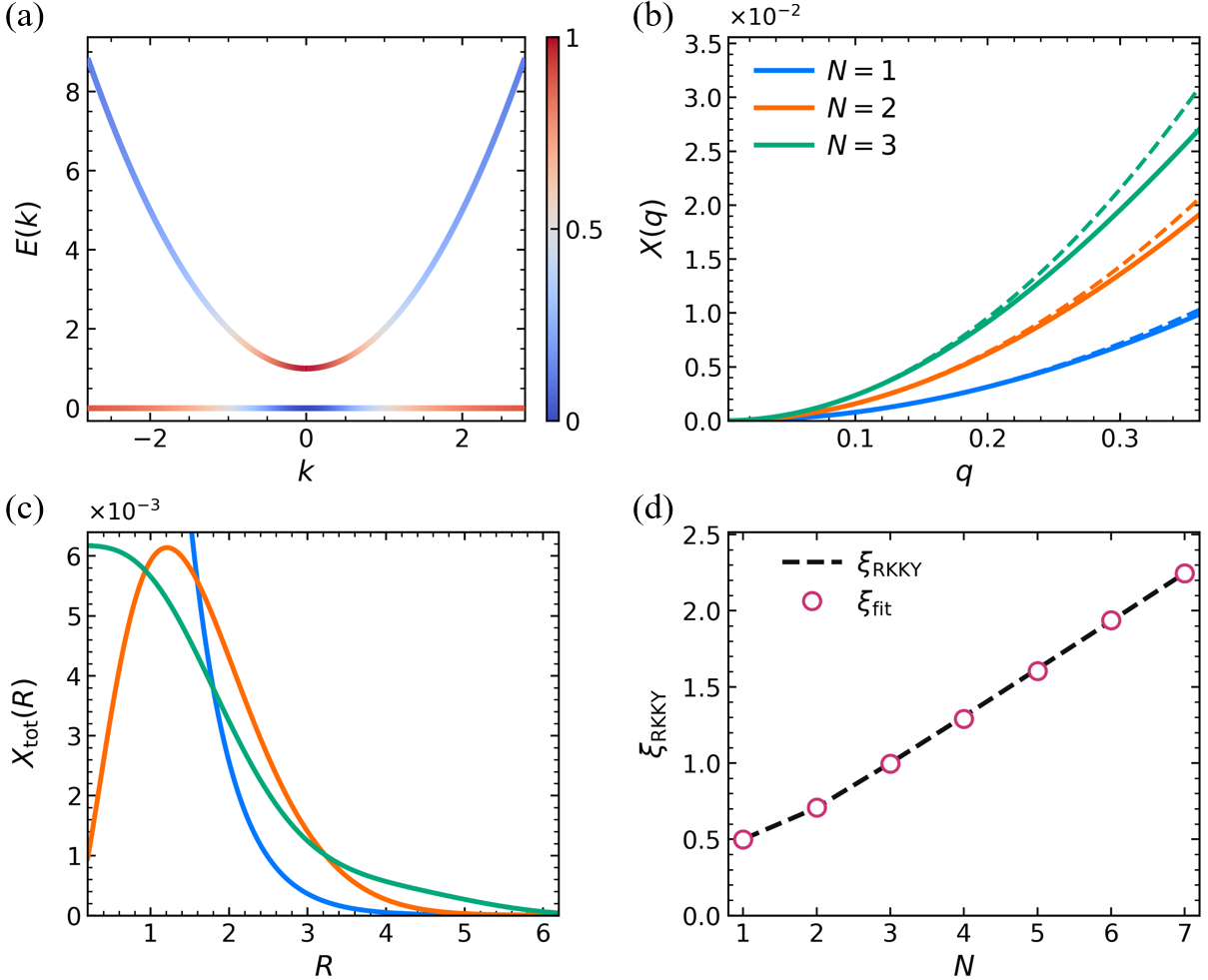}
\caption{Flat-band RKKY interaction. 
(a) Band structure for $N=1$; the color map encodes the orbital weight.
(b) Momentum-space response $X(q)$ for $N=1,2,3$. Solid curves are numerical; dashed curves denote the quantum-metric-length contribution in the small-$q$ limit, $(q l_{\rm QM})^2/(4\pi\Delta_E)$.
(c) Real-space kernels $X_{\rm tot}(R)$, showing exponential decay.
(d) Decay length $\xi_{\rm RKKY}$ versus $N$. The dashed line is the analytic result in Eq.~\eqref{eq:xi_rankone}, and circles are fits to the numerical tails. As expected, increasing $l_{\rm QM}=\sqrt{N}$ leads to a larger $\xiR$.
Parameters are $\lambda=0$ and $a=b=1$.
}
\label{fig:flatband}
\end{figure}

\textit{Ideal Chern flat-band model.---}
We consider an analytically solvable two-band model described by
\begin{align} \label{eq:model}
H(\kk) = |v_N(\kk)\rangle\langle v_N(\kk)| + \lambda k^2 I_2,
\end{align}
with $|v_N(\kk)\rangle = (a k_+^N, b)^T$, where $k_\pm = k_x \pm i k_y$, $N$ is a positive integer, and $a,b$ are real parameters. The $\lambda k^2 I_2$ term introduces a small quadratic dispersion that breaks the exact flatness. For $\lambda=0$, the rank-one structure of Eq.~\eqref{eq:model} enforces a flat band with energy $E_o=0$, separated by a direct gap $\Delta_E = b^2$ from the remote dispersive band with energy $E_u(\kk) = \langle v_N(\kk) \vert v_N(\kk) \rangle = a^2 k^{2N} + \Delta_E$ [Fig.~\ref{fig:flatband}(a)]. The $N=1$ case is realizable in Moir\'e heterostructures with type-II band alignment~\cite{liu2025ideal}. Importantly, this flat band is ideal: it satisfies the quantum-geometric saturation condition $\operatorname{Tr}[g_{\mu\nu}(\kk)] = |\Omega_{xy}(\kk)| = 2N^2 a^2 b^2 k^{2N-2}/E_u^2(\kk)$, thereby carrying Chern number $|C|=N$ and with quantum metric length $l_{\rm QM}=\sqrt{N}$. For $\lambda=0$, the small-$q$ response $X(\qq)$ scales as $X(q)=(q l_{\rm QM})^2/(4\pi \Delta_E)$. The coefficient of this $q^2$ term is determined by the gap-weighted quantum metric (also referred to as the quantum weight~\cite{onishi2025prb}), and its positive sign originates from the positional-shift term~\cite{kitamura2024prl}. We compare the numerical evaluation of $X(q)$ with the analytical results for $N=1,2,3$ in Fig.~\ref{fig:flatband}(b).  
We provide more details in Sec.~A of the Supplementary Material (SM)~\cite{SM}.
However, the RKKY decay length $\xi_{\rm RKKY}$ generally depends on the full $q$ dependence of $X(\qq)$, not merely its small-$q$ expansion; thus $\xi_{\rm RKKY}$ is not determined by $l_{\rm QM}$ alone.

We next examine analytically how quantum geometry and the band gap jointly determine $\xi_{\rm RKKY}$. From Eqs.~\eqref{eq:projector_resolvent} and \eqref{eq:xi-RKKY}, we introduce the band projectors
\begin{align} \label{eq:k-space-proj}
{\cal P}_u(\kk) = \frac{|v_N(\kk)\rangle\langle v_N(\kk)|}{E_u(\kk)}, \;\;
{\cal P}_o(\kk) =  I_2 - {\cal P}_u(\kk).
\end{align}
In real space, these satisfy ${\cal P}_o(\RR) = -{\cal P}_u(\RR)$ for finite $R$, following from the completeness relation ${\cal P}_o(\RR)+{\cal P}_u(\RR)=\delta(\RR)I_2$. Hence both projectors share the same decay length. By the Paley–Wiener theorem~\cite{paley1934fourier}, the Fourier transform of a function analytic in a strip about the real axis decays exponentially, with a rate set by the distance from the real axis to the nearest singularity. The singularities of ${\cal P}_{u/o}(\kk)$ are poles located at the zeros of the analytically continued $E_u(k)$ in the complex plane: $k_m = (b/a)^{1/N} e^{i(\pi+2\pi m)/(2N)}$ with $m=0,1,\dots,2N-1$. The closest poles to the real axis are $k_0$ (and $k_{N-1}$ for $N\ge2$), yielding the decay length $\xi_o = 1/\text{Im}[k_0] = 1/[(b/a)^{1/N}\sin(\pi/(2N))]$ for ${\cal P}_o(\RR)$; for our model, ${\cal R}_u(\RR)$ gives the same $\xi_u = \xi_o$. 
Further details are provided in Sec.~B of the SM~\cite{SM}. 
From Eq.~\eqref{eq:xi-RKKY}, we then obtain
\begin{align} \label{eq:xi_rankone}
\xiR = \xi_o/2 = [ 2 (b/a)^{1/N} \sin\tfrac{\pi}{2N} ]^{-1}.
\end{align}
It is the central result of this work, showing that $\xiR$ is controlled by both $b/a$ (band dispersion) and $N$ (quantum geometry). The underlying mathematical mechanism is analogous to the embedding-independent length scale of flat bands in Ref.~\cite{Lee2026prl}; however, our pole analysis applies equally to continuum and tight-binding models. The RKKY decay length, however, was not addressed in recent work comparing quantum geometry and band dispersion~\cite{huhaoyu2026prl}. Although the flat-band RKKY exchange decays exponentially, the decay length can still be appreciable for realistic parameters. To illustrate this, we apply Eq.~\eqref{eq:xi_rankone} to the real-material model introduced in Ref.~\cite{liu2025ideal}. In that model, the parameters map to our notation as $a=\sqrt{\alpha}$ and $b=\beta/\sqrt{\alpha}$, which gives $\Delta_E=\beta^2/\alpha$. Taking $\alpha\simeq1$~eV$\cdot$nm$^2$ and $\beta\simeq0.2$~eV$\cdot$nm, we obtain $\Delta E\simeq40$~meV and $\xi_{\rm RKKY}\simeq2.5$~nm---a length scale well within the spatial resolution of spin-polarized scanning tunneling microscopy~\cite{Meier2008,zhou2010strength,Wiesendanger2009rmp,choi2019rmp}.

We then verify the above analytical results for the special case $a=b=1$, where the effect of band dispersion is effectively ``eliminated''. For large $N$, the decay length becomes $\xiR = [2\sin(\pi/2N)]^{-1} \approx N/\pi$. Figure~\ref{fig:flatband}(c) shows the numerical results of $X(\RR)$ for $N=1,2,3$, which exhibit exponential decay in the tails. Fitting the decay lengths from these tails [red circles in Fig.~\ref{fig:flatband}(d)] yields excellent agreement with our analytical prediction. In this flat-band limit, $\xiR$ grows with stronger quantum-geometric effects (i.e., larger $N$), highlighting the role of quantum geometry in setting the RKKY decay length. This might suggest that stronger quantum-geometric effects always extend the exchange range; we now show that this is not generally the case.

\begin{figure}[t]
\centering
\includegraphics[width=\columnwidth]{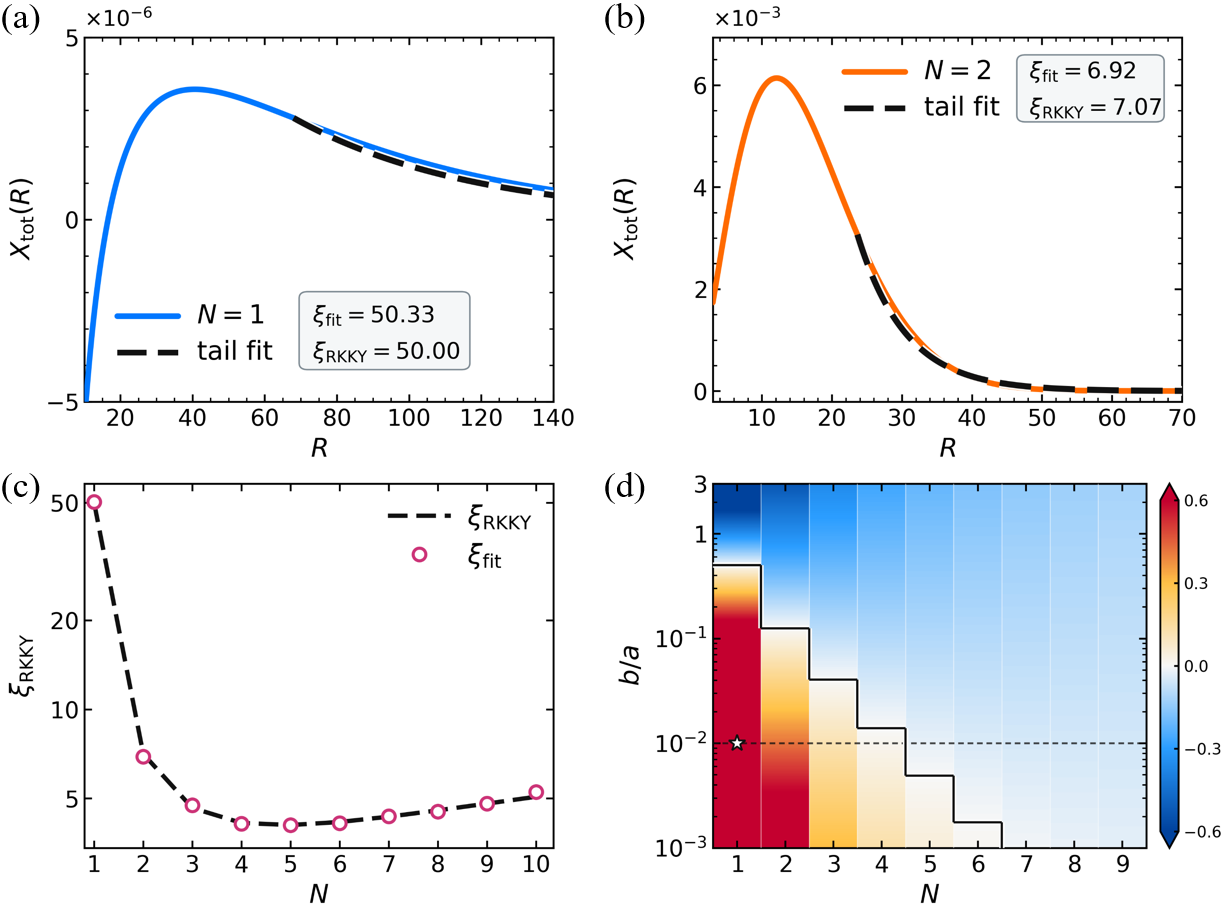}
\caption{Nonmonotonic flat-band RKKY exchange range versus quantum geometry.
(a,b) Real-space kernels for $N=1$ and $N=2$. Despite its larger Chern number, the $N=2$ flat band yields a shorter RKKY range.
(c) Decay length $\xi_{\rm RKKY}$ versus $N$, comparing the analytic formula with numerical fits. The results reveal an unexpected nonmonotonic dependence on $N$ (with $l_{\rm QM}=\sqrt{N}$).
(d) Phase diagram in the $(N,b/a)$ plane. The color map shows $\ln[\xi_{\rm RKKY}(N)/\xi_{\rm RKKY}(N+1)]$, with the black staircase marking $\xiR(N)=\xiR(N+1)$. Red regions indicate where increasing $N$ shortens $\xiR$. The star corresponds to panels (a) and (b).
Parameters are $a=1$ and $b=0.01$.
}
\label{fig:range}
\end{figure}

In fact, $\xiR$ is governed by the competition between the band gap $\Delta_E=b^2$ (taking $a=1$, for instance) and quantum geometry: $\xiR$ grows as $b$ decreases and $N$ increases. Remarkably, however, for a fixed gap, this competition can make $\xiR$ nonmonotonic in $N$, because the prefactor $(b/a)^{1/N}$ in Eq.~\eqref{eq:xi_rankone} increases with $N$ when $b/a < 1$ and can overcome the decrease from the sine factor $\sin(\pi/2N)$. Specifically, $\xiR(N+1)<\xiR(N)$ whenever
\begin{align} \label{eq:winding_anomaly}
\frac{b}{a} < [ \sin\frac{\pi}{2(N+1)} /
\sin\frac{\pi}{2N} ]^{N(N+1)} .    
\end{align}
so that, e.g., $b/a<1/2$ already yields $\xiR(N=1)>\xiR(N=2)$. This is confirmed numerically in Figs.~\ref{fig:range}(a) and \ref{fig:range}(b) for $b/a=0.01$, where the fitted decay lengths are approximately $50$ ($N=1$) and $7$ ($N=2$). The full $N$ dependence at this small gap, shown in Fig.~\ref{fig:range}(c), is nonmonotonic---an initial decrease followed by an increase. Moreover, the phase diagram [Fig.~\ref{fig:range}(d)] maps the $(N,b/a)$ plane, with the color scale $\ln[\xi_{\rm RKKY}(N)/\xi_{\rm RKKY}(N+1)]$; red regions indicate where increasing $N$ shortens the range, bounded by the black staircase. The resulting insight is counterintuitive: quantum geometry does not universally enhance the RKKY range---a flat band with larger Chern number, and thus larger quantum metric weight, can exhibit a shorter exchange range when the gap is sufficiently small, in direct violation of the single geometric scale paradigm. Conversely, flat-band systems with small $l_{\rm QM}$ (small $N$) can still achieve a sizable exchange range by reducing the band gap, e.g., via moir\'e engineering~\cite{liu2025ideal}.

Our flat-band RKKY framework, rooted in pole analysis of projectors, generalizes straightforwardly to tight-binding Chern band models, including the flattened Qi--Wu--Zhang model~\cite{qiwuzhang2006prb}. The Chern flat band can exhibit the same scaling as the $\mathbf{k}\cdot\mathbf{p}$ model: larger $l_{\rm QM}$ reduces $\xiR$ [see Sec.~C of the SM~\cite{SM}].

\begin{figure}[t]
\centering
\includegraphics[width=\columnwidth]{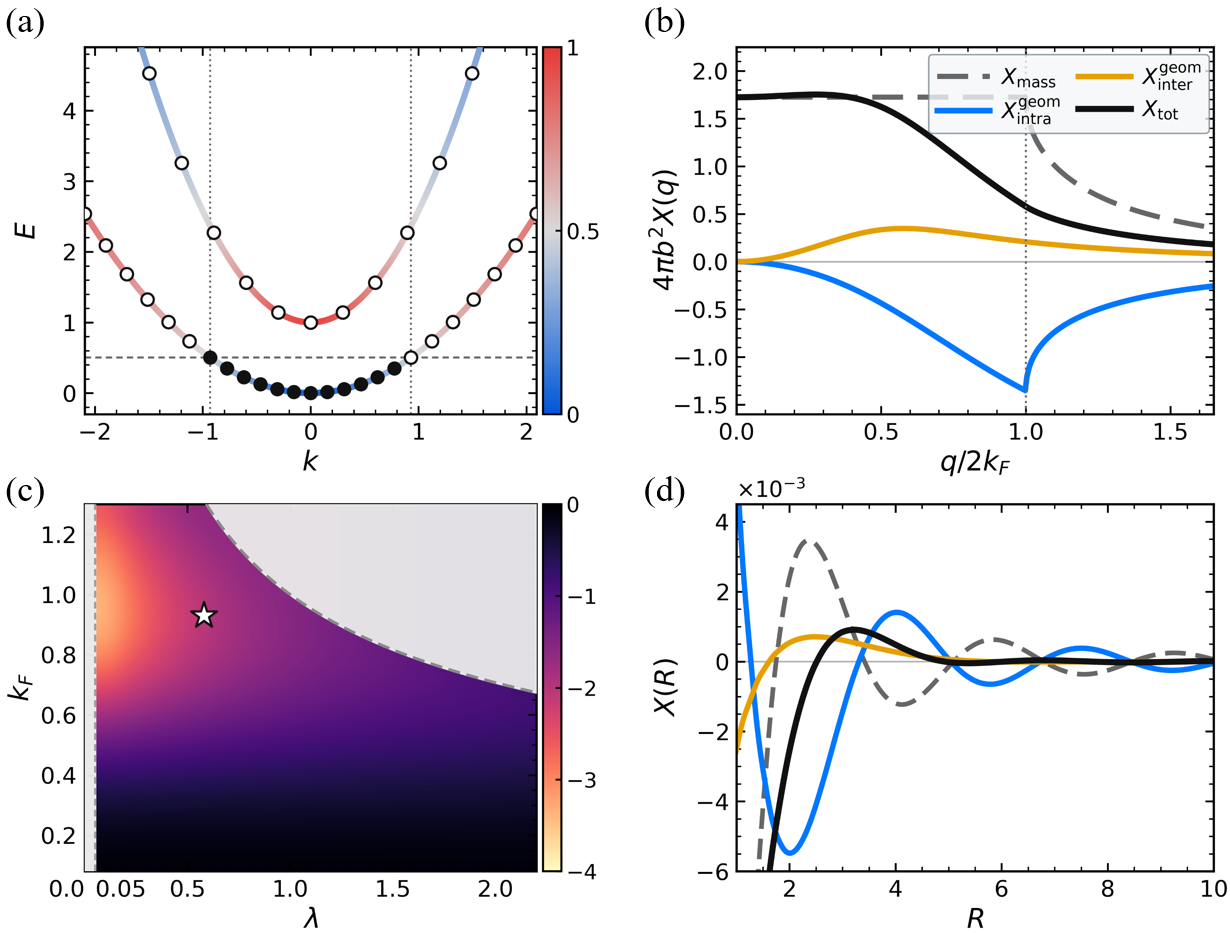}
\caption{Geometric effect on RKKY in the dispersive metal.
(a) Band structure; dashed line: $E_F$; filled (open) circles: occupied (unoccupied) states.
(b) Decomposition of $X_{\rm tot}(q)$ into mass (gray), intraband geometric (blue), interband geometric (orange), and total (black) channels. The intraband channels have a singularity at $q=2k_F$.
(c) Geometric fraction $\eta_{\rm geom}^{2k_F}$ in the $(\lambda,k_F)$ plane for $\lambda>0$ and $E_F<\Delta_E$. Larger values indicate a stronger geometric contribution to RKKY.
(d) Real-space kernel $X(R)$ corresponding to the decomposition in (b). The mass and intraband geometric terms share the same $\pi/k_F$ periodicity but opposite signs.
Parameters are $N=1$, $\lambda=0.58$ and $E_F=0.5$.
}
\label{fig:dispersive}
\end{figure}

\textit{From flat to dispersive bands.---}
In real materials, band flatness is never exact. To explore the crossover from flat to dispersive bands, we turn on the $\lambda$ term in Eq.~\eqref{eq:model}. The resulting band dispersion leaves the $\mathbf{k}$-space projectors unchanged from Eq.~\eqref{eq:k-space-proj}. While we set $N=1$ here, the conclusions are qualitatively the same for larger $N$. The band dispersions are then $E_o(\mathbf{k}) = \lambda k^2$ and $E_u(\mathbf{k}) = (\lambda + a^2)k^2 + b^2$, which are shown in Fig.~\ref{fig:dispersive}(a) with $a = b = 1$, $\lambda = 0.58$, and $E_F = 0.5$ (i.e.,~$k_F=0.93$). For sufficiently small $\lambda$, the flat-band results obtained in Eq.~\eqref{eq:xi_rankone} hold approximately over a large but finite range of distances. Here, instead, we consider the case where both bands are highly dispersive and investigate the role of quantum geometry on RKKY. The emergence of a Fermi surface restores the $2k_F$ branch point in the response function, converting the flat-band exponential decay into a power-law tail. The response now contains both interband and intraband contributions [Fig.~\ref{fig:dispersive}(a)],
\begin{align}
X_{\rm tot}(\qq)=X_{\rm intra}^{\rm mass}(\qq) + X_{\rm intra}^{\rm geom}(\qq) + X_{\rm inter}^{\rm geom}(\qq),
\end{align}
where $X_{\rm intra}^{\rm mass}$ is the conventional intraband Lindhard response (geometry-independent), $X_{\rm intra}^{\rm geom}$ is the intraband geometry-dependent correction, and $X_{\rm inter}^{\rm geom}$ is the geometry-dependent interband contribution.
Their explicit expressions are given in Sec.~D of the SM~\cite{SM}, and numerical results are presented in Fig.~\ref{fig:dispersive}(b).
To leading order in $q^2$, the coefficients of both $X_{\rm intra}^{\rm geom}$ (blue curve) and $X_{\rm inter}^{\rm geom}$ (orange curve) are determined by the quantum metric and exhibit opposite signs~\cite{kitamura2024prl}. Furthermore, we find that the geometric contributions are comparable in magnitude to the mass response near $2k_F$, implying that the geometric contribution is not confined to the nearly flat-band regime.

To quantify the contribution of quantum geometry to $X_{\rm tot}(\qq)$, we define
\begin{align}
\eta_{\rm geom}^{2k_F} = [X_{\rm intra}^{\rm geom}(2k_F)+X_{\rm inter}^{\rm geom}(2k_F) ] / X_{\rm tot}(2k_F).
\end{align}
When $\eta_{\rm geom}^{2k_F} = 0$, the response reduces to the mass-only limit; a nonzero value signals that geometric contributions can be important. Figure~\ref{fig:dispersive}(c) shows $\eta_{\rm geom}^{2k_F}$ in the $\lambda$--$k_F$ plane for the regime where the upper band is entirely empty (i.e., $E_F < \Delta_E$ and $\lambda > 0$). Throughout this region, $\eta_{\rm geom}^{2k_F} < 0$, indicating that geometry opposes the positive mass response and thus frustrates magnetic ordering at the $2k_F$ wavevector---specifically, antiferromagnetism. Notably, at fixed $\lambda$, $|\eta_{\rm geom}^{2k_F}|$ reaches its maximum at $k_F = 1$, marked by the green dashed line, showing that quantum geometry can play a significant role even for highly dispersive bands.

For the parameters used in Fig.~\ref{fig:dispersive}(b) [marked by the star in Fig.~\ref{fig:dispersive}(c)], we obtain $\eta_{\rm geom}^{2k_F}=-1.98$, corresponding to $X_{\rm geom}/X_{\rm mass}\simeq-0.664$. The real-space $X(\RR)$ is shown in Fig.~\ref{fig:dispersive}(d). Both $X_{\rm intra}^{\rm mass}$ and $X_{\rm intra}^{\rm geom}$ exhibit a periodicity of $\pi/k_F \simeq 3.38$, but with opposite signs. In contrast, $X_{\rm inter}^{\rm geom}$ remains exponentially decaying. While experiments could measure only $X_{\rm tot}$ (e.g.,~via spin-polarized scanning tunneling microscopy~\cite{Meier2008,zhou2010strength,Wiesendanger2009rmp,choi2019rmp}), the geometric contribution may be extracted at large $R$ by fitting to the form $\propto (A_{\rm mass}-A_{\rm geom}) \cos(2k_F R + \phi_0)/{R^d}$, if the geometry-free Lindhard amplitude $A_{\rm mass} = 1/(8\pi^2\lambda)$ can be fixed independently from the measured Fermi surface and dispersion.

\begin{figure}[t]
\centering
\includegraphics[width=\columnwidth]{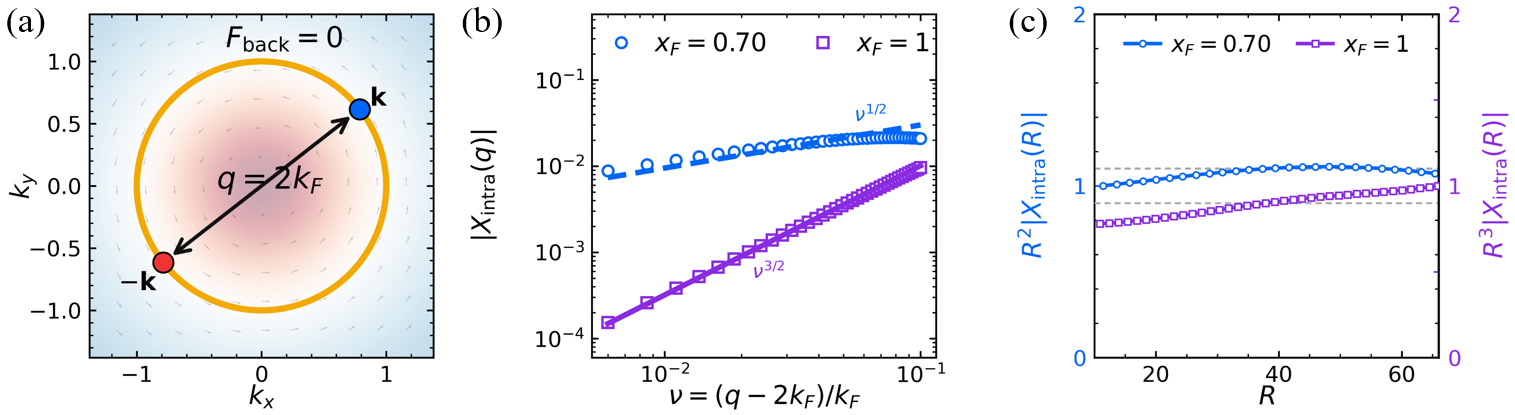}
\caption{Geometric selection of the RKKY power law. 
(a) Antipodal Fermi states connected by $q=2k_F$ at the overlap node $x_F=ak_F/b=1$, where the backscattering matrix element $F_{\rm back}$ vanishes. 
(b) Threshold singularity of $|X_{\rm intra}(q)|$ versus $\nu=(q-2k_F)/k_F$ (analytical corrections subtracted). At $x_F=0.7$: $\nu^{1/2}$ (blue); at $x_F=1$: $\nu^{3/2}$ (purple), with the leading $\nu^{1/2}$ term suppressed.
(c) Compensated real-space tails: $R^{-2}$ for $x_F=0.7$ (blue, left axis) and $R^{-3}$ for $x_F=1$ (purple, right axis). The power-law change is purely geometric and can be tuned by gating $x_F$.
Parameters are $N=1$, $a=b=1$, and $\lambda=0.2$.
}
\label{fig:backscattering}
\end{figure}

\textit{Geometric selection at $2k_F$.---}
As noted above, the most pronounced geometric contribution at $q = 2k_F$ occurs at $k_F = 1$; we now explain this. For both flat and dispersive bands, the geometric contribution originates from the projector overlap $F_{mn}(\kk,\qq)$ entering the kernel in Eq.~\eqref{eq:intro_kernel}. In the dispersive case, the static Kohn anomaly at $|\qq|=2k_F$ is dominated by antipodal pairs on the Fermi circle, $\kk$ and $\kk+\qq=-\kk$ with $|\kk|=k_F$, i.e., by backscattering across the Fermi-surface diameter. For the occupied band, the corresponding backscattering overlap is defined as $F_{\rm back}(k_F) = |\langle u_o(\kk)|u_o(-\kk)\rangle|^2$ at $|\kk|=k_F$, and we find
\begin{align}
F_{\rm back}(k_F) =[(1-x_F^2)/(1+x_F^2)]^2,
\end{align}
which depends only on the dimensionless ratio $x_F=ak_F/b$. Owing to the rotational symmetry of our model, $F_{\rm back}(k_F)$ is isotropic. Figure~\ref{fig:backscattering}(a) illustrates this overlap node at $x_F = 1$, where the antipodal Fermi states are orthogonal ($F_{\rm back}=0$): the exact-backscattering matrix element vanishes. The dashed green line in Fig.~\ref{fig:dispersive}(c) marks this antipodal-backscattering node, $F_{\rm back}(k_F) = 0$. 
This geometric node is a consequence of the inversion operator having a nontrivial matrix representation. For Eq.~\eqref{eq:model}, the odd-$N$ models have ${\cal I} = \tau_z$, while the even-$N$ models have ${\cal I} = \tau_0$. In the odd-$N$ case, $\langle u_o(\kk)|u_o(-\kk)\rangle = \langle u_o(\kk)|\tau_z|u_o(\kk)\rangle = |u_{o,1}(\kk)|^2 - |u_{o,2}(\kk)|^2$, which must vanish due to band inversion. Such a node cannot occur in even-$N$ models.

We now discuss the physical consequence of $F_{\rm back} = 0$ on the power-law tails of RKKY interactions. Since $X_{\rm inter}^{\rm geom}$ is exponentially suppressed at large $R$, it suffices to consider the intraband contribution $X_{\rm intra}(q) = X_{\rm intra}^{\rm mass}(q) + X_{\rm intra}^{\rm geom}(q)$. Near $q \sim 2k_F$, i.e., $\nu \equiv (q-2k_F)/k_F \sim 0$, the non-analytical part of $X_{\rm intra}(q)$ takes the form $X_{\rm intra}(q) \approx \text{Re}[-\rho_0F_{\rm back}(k_F)\nu^{1/2}+A_{3/2}\nu^{3/2}]$, with $\rho_0=1/(4\pi\lambda)$, after neglecting analytical corrections (e.g., $\nu^2$). When $E_F$ is tuned such that $x_F = 1$, the leading-order term vanishes, and the subleading $\nu^{3/2}$ term becomes dominant. This is illustrated in Fig.~\ref{fig:backscattering}(b), which compares $x_F = 0.7$ (blue curve, $\nu^{1/2}$ scaling) and $x_F = 1$ (purple curve, $\nu^{3/2}$ scaling). The latter is enabled entirely by the vanishing of the geometric overlap $F_{\rm back}$, an effect we term \textit{geometric selection}. This threshold selection translates directly into real space: the $\nu^{1/2}$ cusp for $F_{\rm back} \neq 0$ gives the conventional $R^{-2}$ envelope, whereas at the node the leading tail is removed and the envelope becomes $R^{-3}$, with the $2k_F$ oscillation period unchanged [see Sec.~E of the SM~\cite{SM}]. Both cases share the same quadratic dispersion, so the change in power law is purely geometric—analogous to how chirality suppresses backscattering in graphene~\cite{brey2007diluted,sherafati2011doped}, but here controlled by gate tuning via $x_F$.


\textit{Conclusion.---}
In summary, we have shown that quantum geometry drives RKKY exchange from flat to dispersive bands. For a filled flat band, the exchange is purely interband and decays exponentially, with a decay length $\xi_{\rm RKKY}$ determined by the complex-plane poles of the band projectors. This length combines the band gap and quantum geometry into a single scale that can depend non-monotonically on the Chern number—and thus on the quantum metric weight—revealing that stronger quantum geometry can counterintuitively shorten the magnetic exchange range. When dispersion is reintroduced, the exponential tail crosses over to algebraic decay. Remarkably, the nontrivial inversion representation enforces a $1/R^3$ tail instead of the conventional $1/R^2$, without altering the quadratic dispersion—a geometric selection effect accessible by gate tuning. Unlike previous studies of flat-band RKKY (e.g., frustration effects~\cite{bouzerar2021lieb} or Fano-defect models~\cite{luo2024fano}), our work identifies the projector pole structure and the overlap node as the key geometric controls of the exchange range and the asymptotic power law, respectively.

\textit{Acknowledgments.---}
We thank K.-T.~Law, F.-C.~Zhang, C.-X.~Liu, K.-J.~Yang for helpful discussions. This work is supported by National Key R\&D Program of China (Grant No. 2025YFA1411501), the National Natural Science Foundation of China (Grant Nos. 12561160109, 12574148), and the Fundamental Research Funds for the Central Universities (Grant No. 226-2024-00068).

\bibliography{refs2026}

\begin{thebibliography}{49}%
\makeatletter
\providecommand \@ifxundefined [1]{%
 \@ifx{#1\undefined}
}%
\providecommand \@ifnum [1]{%
 \ifnum #1\expandafter \@firstoftwo
 \else \expandafter \@secondoftwo
 \fi
}%
\providecommand \@ifx [1]{%
 \ifx #1\expandafter \@firstoftwo
 \else \expandafter \@secondoftwo
 \fi
}%
\providecommand \natexlab [1]{#1}%
\providecommand \enquote  [1]{``#1''}%
\providecommand \bibnamefont  [1]{#1}%
\providecommand \bibfnamefont [1]{#1}%
\providecommand \citenamefont [1]{#1}%
\providecommand \href@noop [0]{\@secondoftwo}%
\providecommand \href [0]{\begingroup \@sanitize@url \@href}%
\providecommand \@href[1]{\@@startlink{#1}\@@href}%
\providecommand \@@href[1]{\endgroup#1\@@endlink}%
\providecommand \@sanitize@url [0]{\catcode `\\12\catcode `\$12\catcode
  `\&12\catcode `\#12\catcode `\^12\catcode `\_12\catcode `\%12\relax}%
\providecommand \@@startlink[1]{}%
\providecommand \@@endlink[0]{}%
\providecommand \url  [0]{\begingroup\@sanitize@url \@url }%
\providecommand \@url [1]{\endgroup\@href {#1}{\urlprefix }}%
\providecommand \urlprefix  [0]{URL }%
\providecommand \Eprint [0]{\href }%
\providecommand \doibase [0]{https://doi.org/}%
\providecommand \selectlanguage [0]{\@gobble}%
\providecommand \bibinfo  [0]{\@secondoftwo}%
\providecommand \bibfield  [0]{\@secondoftwo}%
\providecommand \translation [1]{[#1]}%
\providecommand \BibitemOpen [0]{}%
\providecommand \bibitemStop [0]{}%
\providecommand \bibitemNoStop [0]{.\EOS\space}%
\providecommand \EOS [0]{\spacefactor3000\relax}%
\providecommand \BibitemShut  [1]{\csname bibitem#1\endcsname}%
\let\auto@bib@innerbib\@empty
\bibitem [{\citenamefont {Mielke}\ and\ \citenamefont
  {Tasaki}(1993)}]{mielke1993ferromagnetism}%
  \BibitemOpen
  \bibfield  {author} {\bibinfo {author} {\bibfnamefont {A.}~\bibnamefont
  {Mielke}}\ and\ \bibinfo {author} {\bibfnamefont {H.}~\bibnamefont
  {Tasaki}},\ }\bibfield  {title} {\bibinfo {title} {Ferromagnetism in the
  hubbard model: Examples from models with degenerate single-electron ground
  states},\ }\href {https://doi.org/10.1007/BF02108079} {\bibfield  {journal}
  {\bibinfo  {journal} {Communications in mathematical physics}\ }\textbf
  {\bibinfo {volume} {158}},\ \bibinfo {pages} {341} (\bibinfo {year}
  {1993})}\BibitemShut {NoStop}%
\bibitem [{\citenamefont {Bistritzer}\ and\ \citenamefont
  {MacDonald}(2011)}]{bistritzer2011moire}%
  \BibitemOpen
  \bibfield  {author} {\bibinfo {author} {\bibfnamefont {R.}~\bibnamefont
  {Bistritzer}}\ and\ \bibinfo {author} {\bibfnamefont {A.~H.}\ \bibnamefont
  {MacDonald}},\ }\bibfield  {title} {\bibinfo {title} {Moir{\'e} bands in
  twisted double-layer graphene},\ }\href
  {https://doi.org/10.1073/pnas.1108174108} {\bibfield  {journal} {\bibinfo
  {journal} {Proceedings of the National Academy of Sciences}\ }\textbf
  {\bibinfo {volume} {108}},\ \bibinfo {pages} {12233} (\bibinfo {year}
  {2011})}\BibitemShut {NoStop}%
\bibitem [{\citenamefont {Cao}\ \emph {et~al.}(2018)\citenamefont {Cao},
  \citenamefont {Fatemi}, \citenamefont {Fang}, \citenamefont {Watanabe},
  \citenamefont {Taniguchi}, \citenamefont {Kaxiras},\ and\ \citenamefont
  {Jarillo-Herrero}}]{cao2018unconventional}%
  \BibitemOpen
  \bibfield  {author} {\bibinfo {author} {\bibfnamefont {Y.}~\bibnamefont
  {Cao}}, \bibinfo {author} {\bibfnamefont {V.}~\bibnamefont {Fatemi}},
  \bibinfo {author} {\bibfnamefont {S.}~\bibnamefont {Fang}}, \bibinfo {author}
  {\bibfnamefont {K.}~\bibnamefont {Watanabe}}, \bibinfo {author}
  {\bibfnamefont {T.}~\bibnamefont {Taniguchi}}, \bibinfo {author}
  {\bibfnamefont {E.}~\bibnamefont {Kaxiras}},\ and\ \bibinfo {author}
  {\bibfnamefont {P.}~\bibnamefont {Jarillo-Herrero}},\ }\bibfield  {title}
  {\bibinfo {title} {Unconventional superconductivity in magic-angle graphene
  superlattices},\ }\href {https://doi.org/10.1038/nature26160} {\bibfield
  {journal} {\bibinfo  {journal} {Nature}\ }\textbf {\bibinfo {volume} {556}},\
  \bibinfo {pages} {43} (\bibinfo {year} {2018})}\BibitemShut {NoStop}%
\bibitem [{\citenamefont {Wu}\ \emph {et~al.}(2018)\citenamefont {Wu},
  \citenamefont {Lovorn}, \citenamefont {Tutuc},\ and\ \citenamefont
  {MacDonald}}]{wu2019prl}%
  \BibitemOpen
  \bibfield  {author} {\bibinfo {author} {\bibfnamefont {F.}~\bibnamefont
  {Wu}}, \bibinfo {author} {\bibfnamefont {T.}~\bibnamefont {Lovorn}}, \bibinfo
  {author} {\bibfnamefont {E.}~\bibnamefont {Tutuc}},\ and\ \bibinfo {author}
  {\bibfnamefont {A.~H.}\ \bibnamefont {MacDonald}},\ }\bibfield  {title}
  {\bibinfo {title} {Hubbard model physics in transition metal dichalcogenide
  moir\'e bands},\ }\href
  {https://link.aps.org/doi/10.1103/PhysRevLett.121.026402} {\bibfield
  {journal} {\bibinfo  {journal} {Phys. Rev. Lett.}\ }\textbf {\bibinfo
  {volume} {121}},\ \bibinfo {pages} {026402} (\bibinfo {year}
  {2018})}\BibitemShut {NoStop}%
\bibitem [{\citenamefont {Tang}\ \emph {et~al.}(2020)\citenamefont {Tang},
  \citenamefont {Li}, \citenamefont {Li}, \citenamefont {Xu}, \citenamefont
  {Liu}, \citenamefont {Barmak}, \citenamefont {Watanabe}, \citenamefont
  {Taniguchi}, \citenamefont {MacDonald}, \citenamefont {Shan} \emph
  {et~al.}}]{tang2020simulation}%
  \BibitemOpen
  \bibfield  {author} {\bibinfo {author} {\bibfnamefont {Y.}~\bibnamefont
  {Tang}}, \bibinfo {author} {\bibfnamefont {L.}~\bibnamefont {Li}}, \bibinfo
  {author} {\bibfnamefont {T.}~\bibnamefont {Li}}, \bibinfo {author}
  {\bibfnamefont {Y.}~\bibnamefont {Xu}}, \bibinfo {author} {\bibfnamefont
  {S.}~\bibnamefont {Liu}}, \bibinfo {author} {\bibfnamefont {K.}~\bibnamefont
  {Barmak}}, \bibinfo {author} {\bibfnamefont {K.}~\bibnamefont {Watanabe}},
  \bibinfo {author} {\bibfnamefont {T.}~\bibnamefont {Taniguchi}}, \bibinfo
  {author} {\bibfnamefont {A.~H.}\ \bibnamefont {MacDonald}}, \bibinfo {author}
  {\bibfnamefont {J.}~\bibnamefont {Shan}}, \emph {et~al.},\ }\bibfield
  {title} {\bibinfo {title} {Simulation of hubbard model physics in {WSe2/WS2}
  moir{\'e} superlattices},\ }\href {https://doi.org/10.1038/s41586-020-2085-3}
  {\bibfield  {journal} {\bibinfo  {journal} {Nature}\ }\textbf {\bibinfo
  {volume} {579}},\ \bibinfo {pages} {353} (\bibinfo {year}
  {2020})}\BibitemShut {NoStop}%
\bibitem [{\citenamefont {Regnault}\ \emph {et~al.}(2022)\citenamefont
  {Regnault}, \citenamefont {Xu}, \citenamefont {Li}, \citenamefont {Ma},
  \citenamefont {Jovanovic}, \citenamefont {Yazdani}, \citenamefont {Parkin},
  \citenamefont {Felser}, \citenamefont {Schoop}, \citenamefont {Ong} \emph
  {et~al.}}]{regnault2022catalogue}%
  \BibitemOpen
  \bibfield  {author} {\bibinfo {author} {\bibfnamefont {N.}~\bibnamefont
  {Regnault}}, \bibinfo {author} {\bibfnamefont {Y.}~\bibnamefont {Xu}},
  \bibinfo {author} {\bibfnamefont {M.-R.}\ \bibnamefont {Li}}, \bibinfo
  {author} {\bibfnamefont {D.-S.}\ \bibnamefont {Ma}}, \bibinfo {author}
  {\bibfnamefont {M.}~\bibnamefont {Jovanovic}}, \bibinfo {author}
  {\bibfnamefont {A.}~\bibnamefont {Yazdani}}, \bibinfo {author} {\bibfnamefont
  {S.~S.}\ \bibnamefont {Parkin}}, \bibinfo {author} {\bibfnamefont
  {C.}~\bibnamefont {Felser}}, \bibinfo {author} {\bibfnamefont {L.~M.}\
  \bibnamefont {Schoop}}, \bibinfo {author} {\bibfnamefont {N.~P.}\
  \bibnamefont {Ong}}, \emph {et~al.},\ }\bibfield  {title} {\bibinfo {title}
  {Catalogue of flat-band stoichiometric materials},\ }\href
  {https://doi.org/10.1038/s41586-022-04519-1} {\bibfield  {journal} {\bibinfo
  {journal} {Nature}\ }\textbf {\bibinfo {volume} {603}},\ \bibinfo {pages}
  {824} (\bibinfo {year} {2022})}\BibitemShut {NoStop}%
\bibitem [{\citenamefont {Kopnin}\ \emph {et~al.}(2011)\citenamefont {Kopnin},
  \citenamefont {Heikkil\"a},\ and\ \citenamefont {Volovik}}]{kopnin2011prb}%
  \BibitemOpen
  \bibfield  {author} {\bibinfo {author} {\bibfnamefont {N.~B.}\ \bibnamefont
  {Kopnin}}, \bibinfo {author} {\bibfnamefont {T.~T.}\ \bibnamefont
  {Heikkil\"a}},\ and\ \bibinfo {author} {\bibfnamefont {G.~E.}\ \bibnamefont
  {Volovik}},\ }\bibfield  {title} {\bibinfo {title} {High-temperature surface
  superconductivity in topological flat-band systems},\ }\href
  {https://link.aps.org/doi/10.1103/PhysRevB.83.220503} {\bibfield  {journal}
  {\bibinfo  {journal} {Phys. Rev. B}\ }\textbf {\bibinfo {volume} {83}},\
  \bibinfo {pages} {220503(R)} (\bibinfo {year} {2011})}\BibitemShut {NoStop}%
\bibitem [{\citenamefont {Peotta}\ and\ \citenamefont
  {T{\"o}rm{\"a}}(2015)}]{peotta2015superfluidity}%
  \BibitemOpen
  \bibfield  {author} {\bibinfo {author} {\bibfnamefont {S.}~\bibnamefont
  {Peotta}}\ and\ \bibinfo {author} {\bibfnamefont {P.}~\bibnamefont
  {T{\"o}rm{\"a}}},\ }\bibfield  {title} {\bibinfo {title} {Superfluidity in
  topologically nontrivial flat bands},\ }\href
  {https://doi.org/10.1038/ncomms9944} {\bibfield  {journal} {\bibinfo
  {journal} {Nature communications}\ }\textbf {\bibinfo {volume} {6}},\
  \bibinfo {pages} {8944} (\bibinfo {year} {2015})}\BibitemShut {NoStop}%
\bibitem [{\citenamefont {Julku}\ \emph {et~al.}(2016)\citenamefont {Julku},
  \citenamefont {Peotta}, \citenamefont {Vanhala}, \citenamefont {Kim},\ and\
  \citenamefont {T\"orm\"a}}]{Julku2016prl}%
  \BibitemOpen
  \bibfield  {author} {\bibinfo {author} {\bibfnamefont {A.}~\bibnamefont
  {Julku}}, \bibinfo {author} {\bibfnamefont {S.}~\bibnamefont {Peotta}},
  \bibinfo {author} {\bibfnamefont {T.~I.}\ \bibnamefont {Vanhala}}, \bibinfo
  {author} {\bibfnamefont {D.-H.}\ \bibnamefont {Kim}},\ and\ \bibinfo {author}
  {\bibfnamefont {P.}~\bibnamefont {T\"orm\"a}},\ }\bibfield  {title} {\bibinfo
  {title} {Geometric origin of superfluidity in the lieb-lattice flat band},\
  }\href {https://doi.org/10.1103/PhysRevLett.117.045303} {\bibfield  {journal}
  {\bibinfo  {journal} {Phys. Rev. Lett.}\ }\textbf {\bibinfo {volume} {117}},\
  \bibinfo {pages} {045303} (\bibinfo {year} {2016})}\BibitemShut {NoStop}%
\bibitem [{\citenamefont {Liang}\ \emph {et~al.}(2017)\citenamefont {Liang},
  \citenamefont {Vanhala}, \citenamefont {Peotta}, \citenamefont {Siro},
  \citenamefont {Harju},\ and\ \citenamefont {T\"orm\"a}}]{Liang2017prb}%
  \BibitemOpen
  \bibfield  {author} {\bibinfo {author} {\bibfnamefont {L.}~\bibnamefont
  {Liang}}, \bibinfo {author} {\bibfnamefont {T.~I.}\ \bibnamefont {Vanhala}},
  \bibinfo {author} {\bibfnamefont {S.}~\bibnamefont {Peotta}}, \bibinfo
  {author} {\bibfnamefont {T.}~\bibnamefont {Siro}}, \bibinfo {author}
  {\bibfnamefont {A.}~\bibnamefont {Harju}},\ and\ \bibinfo {author}
  {\bibfnamefont {P.}~\bibnamefont {T\"orm\"a}},\ }\bibfield  {title} {\bibinfo
  {title} {Band geometry, berry curvature, and superfluid weight},\ }\href
  {https://link.aps.org/doi/10.1103/PhysRevB.95.024515} {\bibfield  {journal}
  {\bibinfo  {journal} {Phys. Rev. B}\ }\textbf {\bibinfo {volume} {95}},\
  \bibinfo {pages} {024515} (\bibinfo {year} {2017})}\BibitemShut {NoStop}%
\bibitem [{\citenamefont {Regnault}\ and\ \citenamefont
  {Bernevig}(2011)}]{Regnault2011prx}%
  \BibitemOpen
  \bibfield  {author} {\bibinfo {author} {\bibfnamefont {N.}~\bibnamefont
  {Regnault}}\ and\ \bibinfo {author} {\bibfnamefont {B.~A.}\ \bibnamefont
  {Bernevig}},\ }\bibfield  {title} {\bibinfo {title} {Fractional chern
  insulator},\ }\href {https://link.aps.org/doi/10.1103/PhysRevX.1.021014}
  {\bibfield  {journal} {\bibinfo  {journal} {Phys. Rev. X}\ }\textbf {\bibinfo
  {volume} {1}},\ \bibinfo {pages} {021014} (\bibinfo {year}
  {2011})}\BibitemShut {NoStop}%
\bibitem [{\citenamefont {Neupert}\ \emph {et~al.}(2011)\citenamefont
  {Neupert}, \citenamefont {Santos}, \citenamefont {Chamon},\ and\
  \citenamefont {Mudry}}]{Neupert2011prl}%
  \BibitemOpen
  \bibfield  {author} {\bibinfo {author} {\bibfnamefont {T.}~\bibnamefont
  {Neupert}}, \bibinfo {author} {\bibfnamefont {L.}~\bibnamefont {Santos}},
  \bibinfo {author} {\bibfnamefont {C.}~\bibnamefont {Chamon}},\ and\ \bibinfo
  {author} {\bibfnamefont {C.}~\bibnamefont {Mudry}},\ }\bibfield  {title}
  {\bibinfo {title} {Fractional quantum hall states at zero magnetic field},\
  }\href {https://doi.org/10.1103/PhysRevLett.106.236804} {\bibfield  {journal}
  {\bibinfo  {journal} {Phys. Rev. Lett.}\ }\textbf {\bibinfo {volume} {106}},\
  \bibinfo {pages} {236804} (\bibinfo {year} {2011})}\BibitemShut {NoStop}%
\bibitem [{\citenamefont {Sun}\ \emph {et~al.}(2011)\citenamefont {Sun},
  \citenamefont {Gu}, \citenamefont {Katsura},\ and\ \citenamefont
  {Das~Sarma}}]{Sun2011prl}%
  \BibitemOpen
  \bibfield  {author} {\bibinfo {author} {\bibfnamefont {K.}~\bibnamefont
  {Sun}}, \bibinfo {author} {\bibfnamefont {Z.}~\bibnamefont {Gu}}, \bibinfo
  {author} {\bibfnamefont {H.}~\bibnamefont {Katsura}},\ and\ \bibinfo {author}
  {\bibfnamefont {S.}~\bibnamefont {Das~Sarma}},\ }\bibfield  {title} {\bibinfo
  {title} {Nearly flatbands with nontrivial topology},\ }\href
  {https://doi.org/10.1103/PhysRevLett.106.236803} {\bibfield  {journal}
  {\bibinfo  {journal} {Phys. Rev. Lett.}\ }\textbf {\bibinfo {volume} {106}},\
  \bibinfo {pages} {236803} (\bibinfo {year} {2011})}\BibitemShut {NoStop}%
\bibitem [{\citenamefont {Tang}\ \emph {et~al.}(2011)\citenamefont {Tang},
  \citenamefont {Mei},\ and\ \citenamefont {Wen}}]{tang2011prl}%
  \BibitemOpen
  \bibfield  {author} {\bibinfo {author} {\bibfnamefont {E.}~\bibnamefont
  {Tang}}, \bibinfo {author} {\bibfnamefont {J.-W.}\ \bibnamefont {Mei}},\ and\
  \bibinfo {author} {\bibfnamefont {X.-G.}\ \bibnamefont {Wen}},\ }\bibfield
  {title} {\bibinfo {title} {High-temperature fractional quantum hall states},\
  }\href {https://doi.org/10.1103/PhysRevLett.106.236802} {\bibfield  {journal}
  {\bibinfo  {journal} {Phys. Rev. Lett.}\ }\textbf {\bibinfo {volume} {106}},\
  \bibinfo {pages} {236802} (\bibinfo {year} {2011})}\BibitemShut {NoStop}%
\bibitem [{\citenamefont {Sheng}\ \emph {et~al.}(2011)\citenamefont {Sheng},
  \citenamefont {Gu}, \citenamefont {Sun},\ and\ \citenamefont
  {Sheng}}]{sheng2011fractional}%
  \BibitemOpen
  \bibfield  {author} {\bibinfo {author} {\bibfnamefont {D.}~\bibnamefont
  {Sheng}}, \bibinfo {author} {\bibfnamefont {Z.-C.}\ \bibnamefont {Gu}},
  \bibinfo {author} {\bibfnamefont {K.}~\bibnamefont {Sun}},\ and\ \bibinfo
  {author} {\bibfnamefont {L.}~\bibnamefont {Sheng}},\ }\bibfield  {title}
  {\bibinfo {title} {Fractional quantum hall effect in the absence of landau
  levels},\ }\href {https://doi.org/10.1038/ncomms1380} {\bibfield  {journal}
  {\bibinfo  {journal} {Nature communications}\ }\textbf {\bibinfo {volume}
  {2}},\ \bibinfo {pages} {389} (\bibinfo {year} {2011})}\BibitemShut {NoStop}%
\bibitem [{\citenamefont {Wu}\ \emph {et~al.}(2007)\citenamefont {Wu},
  \citenamefont {Bergman}, \citenamefont {Balents},\ and\ \citenamefont
  {Das~Sarma}}]{wu2007prl}%
  \BibitemOpen
  \bibfield  {author} {\bibinfo {author} {\bibfnamefont {C.}~\bibnamefont
  {Wu}}, \bibinfo {author} {\bibfnamefont {D.}~\bibnamefont {Bergman}},
  \bibinfo {author} {\bibfnamefont {L.}~\bibnamefont {Balents}},\ and\ \bibinfo
  {author} {\bibfnamefont {S.}~\bibnamefont {Das~Sarma}},\ }\bibfield  {title}
  {\bibinfo {title} {Flat bands and wigner crystallization in the honeycomb
  optical lattice},\ }\href
  {https://link.aps.org/doi/10.1103/PhysRevLett.99.070401} {\bibfield
  {journal} {\bibinfo  {journal} {Phys. Rev. Lett.}\ }\textbf {\bibinfo
  {volume} {99}},\ \bibinfo {pages} {070401} (\bibinfo {year}
  {2007})}\BibitemShut {NoStop}%
\bibitem [{\citenamefont {T\"orm\"a}(2023)}]{paivi2023prl}%
  \BibitemOpen
  \bibfield  {author} {\bibinfo {author} {\bibfnamefont {P.}~\bibnamefont
  {T\"orm\"a}},\ }\bibfield  {title} {\bibinfo {title} {Essay: Where can
  quantum geometry lead us?},\ }\href
  {https://link.aps.org/doi/10.1103/PhysRevLett.131.240001} {\bibfield
  {journal} {\bibinfo  {journal} {Phys. Rev. Lett.}\ }\textbf {\bibinfo
  {volume} {131}},\ \bibinfo {pages} {240001} (\bibinfo {year}
  {2023})}\BibitemShut {NoStop}%
\bibitem [{\citenamefont {Yu}\ \emph {et~al.}(2025)\citenamefont {Yu},
  \citenamefont {Bernevig}, \citenamefont {Queiroz}, \citenamefont {Rossi},
  \citenamefont {T{\"o}rm{\"a}},\ and\ \citenamefont {Yang}}]{yu2025quantum}%
  \BibitemOpen
  \bibfield  {author} {\bibinfo {author} {\bibfnamefont {J.}~\bibnamefont
  {Yu}}, \bibinfo {author} {\bibfnamefont {B.~A.}\ \bibnamefont {Bernevig}},
  \bibinfo {author} {\bibfnamefont {R.}~\bibnamefont {Queiroz}}, \bibinfo
  {author} {\bibfnamefont {E.}~\bibnamefont {Rossi}}, \bibinfo {author}
  {\bibfnamefont {P.}~\bibnamefont {T{\"o}rm{\"a}}},\ and\ \bibinfo {author}
  {\bibfnamefont {B.-J.}\ \bibnamefont {Yang}},\ }\bibfield  {title} {\bibinfo
  {title} {Quantum geometry in quantum materials},\ }\href
  {https://doi.org/10.1038/s41535-025-00801-3} {\bibfield  {journal} {\bibinfo
  {journal} {npj Quantum Materials}\ }\textbf {\bibinfo {volume} {10}},\
  \bibinfo {pages} {101} (\bibinfo {year} {2025})}\BibitemShut {NoStop}%
\bibitem [{\citenamefont {Gao}\ \emph {et~al.}(2025)\citenamefont {Gao},
  \citenamefont {Nagaosa}, \citenamefont {Ni},\ and\ \citenamefont
  {Xu}}]{gao2025quantum}%
  \BibitemOpen
  \bibfield  {author} {\bibinfo {author} {\bibfnamefont {A.}~\bibnamefont
  {Gao}}, \bibinfo {author} {\bibfnamefont {N.}~\bibnamefont {Nagaosa}},
  \bibinfo {author} {\bibfnamefont {N.}~\bibnamefont {Ni}},\ and\ \bibinfo
  {author} {\bibfnamefont {S.-Y.}\ \bibnamefont {Xu}},\ }\bibfield  {title}
  {\bibinfo {title} {Quantum geometry phenomena in condensed matter systems},\
  }\href {https://doi.org/10.48550/arXiv.2508.00469} {\bibfield  {journal}
  {\bibinfo  {journal} {arXiv preprint arXiv:2508.00469}\ } (\bibinfo {year}
  {2025})}\BibitemShut {NoStop}%
\bibitem [{\citenamefont {Liu}\ \emph {et~al.}(2025{\natexlab{a}})\citenamefont
  {Liu}, \citenamefont {Qiang}, \citenamefont {Lu},\ and\ \citenamefont
  {Xie}}]{liu2025quantum}%
  \BibitemOpen
  \bibfield  {author} {\bibinfo {author} {\bibfnamefont {T.}~\bibnamefont
  {Liu}}, \bibinfo {author} {\bibfnamefont {X.-B.}\ \bibnamefont {Qiang}},
  \bibinfo {author} {\bibfnamefont {H.-Z.}\ \bibnamefont {Lu}},\ and\ \bibinfo
  {author} {\bibfnamefont {X.}~\bibnamefont {Xie}},\ }\bibfield  {title}
  {\bibinfo {title} {Quantum geometry in condensed matter},\ }\href
  {https://doi.org/10.1093/nsr/nwae334} {\bibfield  {journal} {\bibinfo
  {journal} {National Science Review}\ }\textbf {\bibinfo {volume} {12}},\
  \bibinfo {pages} {nwae334} (\bibinfo {year}
  {2025}{\natexlab{a}})}\BibitemShut {NoStop}%
\bibitem [{\citenamefont {Jiang}\ \emph {et~al.}(2025)\citenamefont {Jiang},
  \citenamefont {Holder},\ and\ \citenamefont {Yan}}]{jiang2025revealing}%
  \BibitemOpen
  \bibfield  {author} {\bibinfo {author} {\bibfnamefont {Y.}~\bibnamefont
  {Jiang}}, \bibinfo {author} {\bibfnamefont {T.}~\bibnamefont {Holder}},\ and\
  \bibinfo {author} {\bibfnamefont {B.}~\bibnamefont {Yan}},\ }\bibfield
  {title} {\bibinfo {title} {Revealing quantum geometry in nonlinear quantum
  materials},\ }\href {https://doi.org/10.1088/1361-6633/ade454} {\bibfield
  {journal} {\bibinfo  {journal} {Reports on Progress in Physics}\ }\textbf
  {\bibinfo {volume} {88}},\ \bibinfo {pages} {076502} (\bibinfo {year}
  {2025})}\BibitemShut {NoStop}%
\bibitem [{\citenamefont {Verma}\ \emph {et~al.}(2026)\citenamefont {Verma},
  \citenamefont {Moll}, \citenamefont {Holder},\ and\ \citenamefont
  {Queiroz}}]{verma2026quantum}%
  \BibitemOpen
  \bibfield  {author} {\bibinfo {author} {\bibfnamefont {N.}~\bibnamefont
  {Verma}}, \bibinfo {author} {\bibfnamefont {P.~J.}\ \bibnamefont {Moll}},
  \bibinfo {author} {\bibfnamefont {T.}~\bibnamefont {Holder}},\ and\ \bibinfo
  {author} {\bibfnamefont {R.}~\bibnamefont {Queiroz}},\ }\bibfield  {title}
  {\bibinfo {title} {Quantum geometry and the hidden scales in materials},\
  }\href {https://doi.org/10.1038/s42254-026-00923-y} {\bibfield  {journal}
  {\bibinfo  {journal} {Nature Reviews Physics}\ }\textbf {\bibinfo {volume}
  {8}},\ \bibinfo {pages} {226–239} (\bibinfo {year} {2026})}\BibitemShut
  {NoStop}%
\bibitem [{\citenamefont {Kitamura}\ \emph {et~al.}(2026)\citenamefont
  {Kitamura}, \citenamefont {Daido},\ and\ \citenamefont
  {Yanase}}]{kitamura2026quantum}%
  \BibitemOpen
  \bibfield  {author} {\bibinfo {author} {\bibfnamefont {T.}~\bibnamefont
  {Kitamura}}, \bibinfo {author} {\bibfnamefont {A.}~\bibnamefont {Daido}},\
  and\ \bibinfo {author} {\bibfnamefont {Y.}~\bibnamefont {Yanase}},\
  }\bibfield  {title} {\bibinfo {title} {Quantum geometry in correlated
  electron phases: From flatband to dispersive band},\ }\href
  {https://doi.org/10.1063/5.0319517} {\bibfield  {journal} {\bibinfo
  {journal} {Applied Physics Letters}\ }\textbf {\bibinfo {volume} {128}}
  (\bibinfo {year} {2026})}\BibitemShut {NoStop}%
\bibitem [{\citenamefont {Huhtinen}\ \emph {et~al.}(2022)\citenamefont
  {Huhtinen}, \citenamefont {Herzog-Arbeitman}, \citenamefont {Chew},
  \citenamefont {Bernevig},\ and\ \citenamefont {T\"orm\"a}}]{Kukka2022prb}%
  \BibitemOpen
  \bibfield  {author} {\bibinfo {author} {\bibfnamefont {K.-E.}\ \bibnamefont
  {Huhtinen}}, \bibinfo {author} {\bibfnamefont {J.}~\bibnamefont
  {Herzog-Arbeitman}}, \bibinfo {author} {\bibfnamefont {A.}~\bibnamefont
  {Chew}}, \bibinfo {author} {\bibfnamefont {B.~A.}\ \bibnamefont {Bernevig}},\
  and\ \bibinfo {author} {\bibfnamefont {P.}~\bibnamefont {T\"orm\"a}},\
  }\bibfield  {title} {\bibinfo {title} {Revisiting flat band
  superconductivity: Dependence on minimal quantum metric and band touchings},\
  }\href {https://link.aps.org/doi/10.1103/PhysRevB.106.014518} {\bibfield
  {journal} {\bibinfo  {journal} {Phys. Rev. B}\ }\textbf {\bibinfo {volume}
  {106}},\ \bibinfo {pages} {014518} (\bibinfo {year} {2022})}\BibitemShut
  {NoStop}%
\bibitem [{\citenamefont {Tian}\ \emph {et~al.}(2023)\citenamefont {Tian},
  \citenamefont {Gao}, \citenamefont {Zhang}, \citenamefont {Che},
  \citenamefont {Xu}, \citenamefont {Cheung}, \citenamefont {Watanabe},
  \citenamefont {Taniguchi}, \citenamefont {Randeria}, \citenamefont {Zhang}
  \emph {et~al.}}]{tian2023evidence}%
  \BibitemOpen
  \bibfield  {author} {\bibinfo {author} {\bibfnamefont {H.}~\bibnamefont
  {Tian}}, \bibinfo {author} {\bibfnamefont {X.}~\bibnamefont {Gao}}, \bibinfo
  {author} {\bibfnamefont {Y.}~\bibnamefont {Zhang}}, \bibinfo {author}
  {\bibfnamefont {S.}~\bibnamefont {Che}}, \bibinfo {author} {\bibfnamefont
  {T.}~\bibnamefont {Xu}}, \bibinfo {author} {\bibfnamefont {P.}~\bibnamefont
  {Cheung}}, \bibinfo {author} {\bibfnamefont {K.}~\bibnamefont {Watanabe}},
  \bibinfo {author} {\bibfnamefont {T.}~\bibnamefont {Taniguchi}}, \bibinfo
  {author} {\bibfnamefont {M.}~\bibnamefont {Randeria}}, \bibinfo {author}
  {\bibfnamefont {F.}~\bibnamefont {Zhang}}, \emph {et~al.},\ }\bibfield
  {title} {\bibinfo {title} {Evidence for dirac flat band superconductivity
  enabled by quantum geometry},\ }\href
  {https://doi.org/10.1038/s41586-022-05576-2} {\bibfield  {journal} {\bibinfo
  {journal} {Nature}\ }\textbf {\bibinfo {volume} {614}},\ \bibinfo {pages}
  {440} (\bibinfo {year} {2023})}\BibitemShut {NoStop}%
\bibitem [{\citenamefont {Tanaka}\ \emph {et~al.}(2025)\citenamefont {Tanaka},
  \citenamefont {Wang}, \citenamefont {Dinh}, \citenamefont {Rodan-Legrain},
  \citenamefont {Zaman}, \citenamefont {Hays}, \citenamefont {Almanakly},
  \citenamefont {Kannan}, \citenamefont {Kim}, \citenamefont {Niedzielski}
  \emph {et~al.}}]{tanaka2025superfluid}%
  \BibitemOpen
  \bibfield  {author} {\bibinfo {author} {\bibfnamefont {M.}~\bibnamefont
  {Tanaka}}, \bibinfo {author} {\bibfnamefont {J.~{\^I}.-J.}\ \bibnamefont
  {Wang}}, \bibinfo {author} {\bibfnamefont {T.~H.}\ \bibnamefont {Dinh}},
  \bibinfo {author} {\bibfnamefont {D.}~\bibnamefont {Rodan-Legrain}}, \bibinfo
  {author} {\bibfnamefont {S.}~\bibnamefont {Zaman}}, \bibinfo {author}
  {\bibfnamefont {M.}~\bibnamefont {Hays}}, \bibinfo {author} {\bibfnamefont
  {A.}~\bibnamefont {Almanakly}}, \bibinfo {author} {\bibfnamefont
  {B.}~\bibnamefont {Kannan}}, \bibinfo {author} {\bibfnamefont {D.~K.}\
  \bibnamefont {Kim}}, \bibinfo {author} {\bibfnamefont {B.~M.}\ \bibnamefont
  {Niedzielski}}, \emph {et~al.},\ }\bibfield  {title} {\bibinfo {title}
  {Superfluid stiffness of magic-angle twisted bilayer graphene},\ }\href
  {https://doi.org/10.1038/s41586-024-08494-7} {\bibfield  {journal} {\bibinfo
  {journal} {Nature}\ }\textbf {\bibinfo {volume} {638}},\ \bibinfo {pages}
  {99} (\bibinfo {year} {2025})}\BibitemShut {NoStop}%
\bibitem [{\citenamefont {Chen}\ and\ \citenamefont {Law}(2024)}]{Chen2024prl}%
  \BibitemOpen
  \bibfield  {author} {\bibinfo {author} {\bibfnamefont {S.~A.}\ \bibnamefont
  {Chen}}\ and\ \bibinfo {author} {\bibfnamefont {K.~T.}\ \bibnamefont {Law}},\
  }\bibfield  {title} {\bibinfo {title} {Ginzburg-landau theory of flat-band
  superconductors with quantum metric},\ }\href
  {https://doi.org/10.1103/PhysRevLett.132.026002} {\bibfield  {journal}
  {\bibinfo  {journal} {Phys. Rev. Lett.}\ }\textbf {\bibinfo {volume} {132}},\
  \bibinfo {pages} {026002} (\bibinfo {year} {2024})}\BibitemShut {NoStop}%
\bibitem [{\citenamefont {Iskin}(2023)}]{Iskin2023prb}%
  \BibitemOpen
  \bibfield  {author} {\bibinfo {author} {\bibfnamefont {M.}~\bibnamefont
  {Iskin}},\ }\bibfield  {title} {\bibinfo {title} {Extracting
  quantum-geometric effects from ginzburg-landau theory in a multiband hubbard
  model},\ }\href {https://doi.org/10.1103/PhysRevB.107.224505} {\bibfield
  {journal} {\bibinfo  {journal} {Phys. Rev. B}\ }\textbf {\bibinfo {volume}
  {107}},\ \bibinfo {pages} {224505} (\bibinfo {year} {2023})}\BibitemShut
  {NoStop}%
\bibitem [{\citenamefont {Hu}\ \emph {et~al.}(2025)\citenamefont {Hu},
  \citenamefont {Chen},\ and\ \citenamefont {Law}}]{hu2025anomalous}%
  \BibitemOpen
  \bibfield  {author} {\bibinfo {author} {\bibfnamefont {J.-X.}\ \bibnamefont
  {Hu}}, \bibinfo {author} {\bibfnamefont {S.~A.}\ \bibnamefont {Chen}},\ and\
  \bibinfo {author} {\bibfnamefont {K.~T.}\ \bibnamefont {Law}},\ }\bibfield
  {title} {\bibinfo {title} {Anomalous coherence length in superconductors with
  quantum metric},\ }\href {https://doi.org/10.1038/s42005-024-01930-0}
  {\bibfield  {journal} {\bibinfo  {journal} {Communications Physics}\ }\textbf
  {\bibinfo {volume} {8}},\ \bibinfo {pages} {20} (\bibinfo {year}
  {2025})}\BibitemShut {NoStop}%
\bibitem [{\citenamefont {Li}\ \emph {et~al.}(2025)\citenamefont {Li},
  \citenamefont {Zhang},\ and\ \citenamefont {Hu}}]{li2025vortex}%
  \BibitemOpen
  \bibfield  {author} {\bibinfo {author} {\bibfnamefont {C.}~\bibnamefont
  {Li}}, \bibinfo {author} {\bibfnamefont {F.-C.}\ \bibnamefont {Zhang}},\ and\
  \bibinfo {author} {\bibfnamefont {L.-H.}\ \bibnamefont {Hu}},\ }\bibfield
  {title} {\bibinfo {title} {Vortex states and coherence lengths in flat-band
  superconductors},\ }\href {https://doi.org/10.48550/arXiv.2505.01682}
  {\bibfield  {journal} {\bibinfo  {journal} {arXiv preprint arXiv:2505.01682}\
  } (\bibinfo {year} {2025})}\BibitemShut {NoStop}%
\bibitem [{\citenamefont {Marzari}\ and\ \citenamefont
  {Vanderbilt}(1997)}]{Marzari1997prb}%
  \BibitemOpen
  \bibfield  {author} {\bibinfo {author} {\bibfnamefont {N.}~\bibnamefont
  {Marzari}}\ and\ \bibinfo {author} {\bibfnamefont {D.}~\bibnamefont
  {Vanderbilt}},\ }\bibfield  {title} {\bibinfo {title} {Maximally localized
  generalized wannier functions for composite energy bands},\ }\href
  {https://link.aps.org/doi/10.1103/PhysRevB.56.12847} {\bibfield  {journal}
  {\bibinfo  {journal} {Phys. Rev. B}\ }\textbf {\bibinfo {volume} {56}},\
  \bibinfo {pages} {12847} (\bibinfo {year} {1997})}\BibitemShut {NoStop}%
\bibitem [{\citenamefont {Coleman}(2015)}]{coleman2015introduction}%
  \BibitemOpen
  \bibfield  {author} {\bibinfo {author} {\bibfnamefont {P.}~\bibnamefont
  {Coleman}},\ }\href@noop {} {\emph {\bibinfo {title} {Introduction to
  many-body physics}}}\ (\bibinfo  {publisher} {Cambridge University Press},\
  \bibinfo {year} {2015})\BibitemShut {NoStop}%
\bibitem [{\citenamefont {Kato}\ and\ \citenamefont
  {Kat{\aa}o}(1966)}]{kato1966perturbation}%
  \BibitemOpen
  \bibfield  {author} {\bibinfo {author} {\bibfnamefont {T.}~\bibnamefont
  {Kato}}\ and\ \bibinfo {author} {\bibfnamefont {T.}~\bibnamefont
  {Kat{\aa}o}},\ }\href@noop {} {\emph {\bibinfo {title} {Perturbation theory
  for linear operators}}},\ Vol.\ \bibinfo {volume} {132}\ (\bibinfo
  {publisher} {Springer},\ \bibinfo {year} {1966})\BibitemShut {NoStop}%
\bibitem [{\citenamefont {Liu}\ \emph {et~al.}(2025{\natexlab{b}})\citenamefont
  {Liu}, \citenamefont {Aryal}, \citenamefont {Calugaru}, \citenamefont {Fang},
  \citenamefont {Yang}, \citenamefont {Hu}, \citenamefont {Yan}, \citenamefont
  {Bernevig},\ and\ \citenamefont {Liu}}]{liu2025ideal}%
  \BibitemOpen
  \bibfield  {author} {\bibinfo {author} {\bibfnamefont {Y.}~\bibnamefont
  {Liu}}, \bibinfo {author} {\bibfnamefont {A.}~\bibnamefont {Aryal}}, \bibinfo
  {author} {\bibfnamefont {D.}~\bibnamefont {Calugaru}}, \bibinfo {author}
  {\bibfnamefont {Z.}~\bibnamefont {Fang}}, \bibinfo {author} {\bibfnamefont
  {K.}~\bibnamefont {Yang}}, \bibinfo {author} {\bibfnamefont {H.}~\bibnamefont
  {Hu}}, \bibinfo {author} {\bibfnamefont {Q.}~\bibnamefont {Yan}}, \bibinfo
  {author} {\bibfnamefont {B.~A.}\ \bibnamefont {Bernevig}},\ and\ \bibinfo
  {author} {\bibfnamefont {C.-X.}\ \bibnamefont {Liu}},\ }\bibfield  {title}
  {\bibinfo {title} {``ideal'' topological heavy fermion model in
  two-dimensional moir$\backslash$'e heterostructures with type-ii band
  alignment},\ }\href {https://doi.org/10.48550/arXiv.2507.06168} {\bibfield
  {journal} {\bibinfo  {journal} {arXiv preprint arXiv:2507.06168}\ } (\bibinfo
  {year} {2025}{\natexlab{b}})}\BibitemShut {NoStop}%
\bibitem [{\citenamefont {Onishi}\ and\ \citenamefont
  {Fu}(2025)}]{onishi2025prb}%
  \BibitemOpen
  \bibfield  {author} {\bibinfo {author} {\bibfnamefont {Y.}~\bibnamefont
  {Onishi}}\ and\ \bibinfo {author} {\bibfnamefont {L.}~\bibnamefont {Fu}},\
  }\bibfield  {title} {\bibinfo {title} {Quantum weight: A fundamental property
  of quantum many-body systems},\ }\href
  {https://link.aps.org/doi/10.1103/PhysRevResearch.7.023158} {\bibfield
  {journal} {\bibinfo  {journal} {Phys. Rev. Res.}\ }\textbf {\bibinfo {volume}
  {7}},\ \bibinfo {pages} {023158} (\bibinfo {year} {2025})}\BibitemShut
  {NoStop}%
\bibitem [{\citenamefont {Kitamura}\ \emph {et~al.}(2024)\citenamefont
  {Kitamura}, \citenamefont {Daido},\ and\ \citenamefont
  {Yanase}}]{kitamura2024prl}%
  \BibitemOpen
  \bibfield  {author} {\bibinfo {author} {\bibfnamefont {T.}~\bibnamefont
  {Kitamura}}, \bibinfo {author} {\bibfnamefont {A.}~\bibnamefont {Daido}},\
  and\ \bibinfo {author} {\bibfnamefont {Y.}~\bibnamefont {Yanase}},\
  }\bibfield  {title} {\bibinfo {title} {Spin-triplet superconductivity from
  quantum-geometry-induced ferromagnetic fluctuation},\ }\href
  {https://link.aps.org/doi/10.1103/PhysRevLett.132.036001} {\bibfield
  {journal} {\bibinfo  {journal} {Phys. Rev. Lett.}\ }\textbf {\bibinfo
  {volume} {132}},\ \bibinfo {pages} {036001} (\bibinfo {year}
  {2024})}\BibitemShut {NoStop}%
\bibitem [{SM()}]{SM}%
  \BibitemOpen
  \href@noop {} {}\bibinfo {note} {See the Supplemental Material for details on
  the small-$q$ response (Sec.~A), the pole analysis of projects (Sec.~B), the
  tight-binding generalization (Sec.~C), the dispersive RKKY decomposition
  (Sec.~D), and the geometric selection analysis of the power laws
  (Sec.~E).}\BibitemShut {Stop}%
\bibitem [{\citenamefont {Paley}\ and\ \citenamefont
  {Wiener}(1934)}]{paley1934fourier}%
  \BibitemOpen
  \bibfield  {author} {\bibinfo {author} {\bibfnamefont {R.~E. A.~C.}\
  \bibnamefont {Paley}}\ and\ \bibinfo {author} {\bibfnamefont
  {N.}~\bibnamefont {Wiener}},\ }\href@noop {} {\emph {\bibinfo {title}
  {Fourier transforms in the complex domain}}},\ Vol.~\bibinfo {volume} {19}\
  (\bibinfo  {publisher} {American Mathematical Soc.},\ \bibinfo {year}
  {1934})\BibitemShut {NoStop}%
\bibitem [{\citenamefont {Lee}\ \emph {et~al.}(2026)\citenamefont {Lee},
  \citenamefont {Lee},\ and\ \citenamefont {Yang}}]{Lee2026prl}%
  \BibitemOpen
  \bibfield  {author} {\bibinfo {author} {\bibfnamefont {S.}~\bibnamefont
  {Lee}}, \bibinfo {author} {\bibfnamefont {S.~H.}\ \bibnamefont {Lee}},\ and\
  \bibinfo {author} {\bibfnamefont {B.-J.}\ \bibnamefont {Yang}},\ }\bibfield
  {title} {\bibinfo {title} {Embedding independent length scale of flat
  bands},\ }\href {https://doi.org/10.1103/tgtq-nrrt} {\bibfield  {journal}
  {\bibinfo  {journal} {Phys. Rev. Lett.}\ }\textbf {\bibinfo {volume} {137}},\
  \bibinfo {pages} {016401} (\bibinfo {year} {2026})}\BibitemShut {NoStop}%
\bibitem [{\citenamefont {Hu}\ \emph {et~al.}(2026)\citenamefont {Hu},
  \citenamefont {Vafek}, \citenamefont {Haule},\ and\ \citenamefont
  {Bernevig}}]{huhaoyu2026prl}%
  \BibitemOpen
  \bibfield  {author} {\bibinfo {author} {\bibfnamefont {H.}~\bibnamefont
  {Hu}}, \bibinfo {author} {\bibfnamefont {O.}~\bibnamefont {Vafek}}, \bibinfo
  {author} {\bibfnamefont {K.}~\bibnamefont {Haule}},\ and\ \bibinfo {author}
  {\bibfnamefont {B.~A.}\ \bibnamefont {Bernevig}},\ }\bibfield  {title}
  {\bibinfo {title} {Ferromagnetism versus antiferromagnetism in narrow-band
  systems: Competition between quantum geometry and band dispersion},\ }\href
  {https://link.aps.org/doi/10.1103/zdyq-3m9x} {\bibfield  {journal} {\bibinfo
  {journal} {Phys. Rev. Lett.}\ }\textbf {\bibinfo {volume} {136}},\ \bibinfo
  {pages} {256505} (\bibinfo {year} {2026})}\BibitemShut {NoStop}%
\bibitem [{\citenamefont {Meier}\ \emph {et~al.}(2008)\citenamefont {Meier},
  \citenamefont {Zhou}, \citenamefont {Wiebe},\ and\ \citenamefont
  {Wiesendanger}}]{Meier2008}%
  \BibitemOpen
  \bibfield  {author} {\bibinfo {author} {\bibfnamefont {F.}~\bibnamefont
  {Meier}}, \bibinfo {author} {\bibfnamefont {L.}~\bibnamefont {Zhou}},
  \bibinfo {author} {\bibfnamefont {J.}~\bibnamefont {Wiebe}},\ and\ \bibinfo
  {author} {\bibfnamefont {R.}~\bibnamefont {Wiesendanger}},\ }\bibfield
  {title} {\bibinfo {title} {Revealing magnetic interactions from single-atom
  magnetization curves},\ }\href {http://dx.doi.org/10.1126/science.1154415}
  {\bibfield  {journal} {\bibinfo  {journal} {Science}\ }\textbf {\bibinfo
  {volume} {320}},\ \bibinfo {pages} {82–86} (\bibinfo {year}
  {2008})}\BibitemShut {NoStop}%
\bibitem [{\citenamefont {Zhou}\ \emph {et~al.}(2010)\citenamefont {Zhou},
  \citenamefont {Wiebe}, \citenamefont {Lounis}, \citenamefont {Vedmedenko},
  \citenamefont {Meier}, \citenamefont {Bl{\"u}gel}, \citenamefont
  {Dederichs},\ and\ \citenamefont {Wiesendanger}}]{zhou2010strength}%
  \BibitemOpen
  \bibfield  {author} {\bibinfo {author} {\bibfnamefont {L.}~\bibnamefont
  {Zhou}}, \bibinfo {author} {\bibfnamefont {J.}~\bibnamefont {Wiebe}},
  \bibinfo {author} {\bibfnamefont {S.}~\bibnamefont {Lounis}}, \bibinfo
  {author} {\bibfnamefont {E.}~\bibnamefont {Vedmedenko}}, \bibinfo {author}
  {\bibfnamefont {F.}~\bibnamefont {Meier}}, \bibinfo {author} {\bibfnamefont
  {S.}~\bibnamefont {Bl{\"u}gel}}, \bibinfo {author} {\bibfnamefont {P.~H.}\
  \bibnamefont {Dederichs}},\ and\ \bibinfo {author} {\bibfnamefont
  {R.}~\bibnamefont {Wiesendanger}},\ }\bibfield  {title} {\bibinfo {title}
  {Strength and directionality of surface {Ruderman--Kittel--Kasuya--Yosida}
  interaction mapped on the atomic scale},\ }\href
  {https://doi.org/10.1038/nphys1514} {\bibfield  {journal} {\bibinfo
  {journal} {Nature Physics}\ }\textbf {\bibinfo {volume} {6}},\ \bibinfo
  {pages} {187} (\bibinfo {year} {2010})}\BibitemShut {NoStop}%
\bibitem [{\citenamefont {Wiesendanger}(2009)}]{Wiesendanger2009rmp}%
  \BibitemOpen
  \bibfield  {author} {\bibinfo {author} {\bibfnamefont {R.}~\bibnamefont
  {Wiesendanger}},\ }\bibfield  {title} {\bibinfo {title} {Spin mapping at the
  nanoscale and atomic scale},\ }\href
  {https://link.aps.org/doi/10.1103/RevModPhys.81.1495} {\bibfield  {journal}
  {\bibinfo  {journal} {Rev. Mod. Phys.}\ }\textbf {\bibinfo {volume} {81}},\
  \bibinfo {pages} {1495} (\bibinfo {year} {2009})}\BibitemShut {NoStop}%
\bibitem [{\citenamefont {Choi}\ \emph {et~al.}(2019)\citenamefont {Choi},
  \citenamefont {Lorente}, \citenamefont {Wiebe}, \citenamefont {von Bergmann},
  \citenamefont {Otte},\ and\ \citenamefont {Heinrich}}]{choi2019rmp}%
  \BibitemOpen
  \bibfield  {author} {\bibinfo {author} {\bibfnamefont {D.-J.}\ \bibnamefont
  {Choi}}, \bibinfo {author} {\bibfnamefont {N.}~\bibnamefont {Lorente}},
  \bibinfo {author} {\bibfnamefont {J.}~\bibnamefont {Wiebe}}, \bibinfo
  {author} {\bibfnamefont {K.}~\bibnamefont {von Bergmann}}, \bibinfo {author}
  {\bibfnamefont {A.~F.}\ \bibnamefont {Otte}},\ and\ \bibinfo {author}
  {\bibfnamefont {A.~J.}\ \bibnamefont {Heinrich}},\ }\bibfield  {title}
  {\bibinfo {title} {Colloquium: Atomic spin chains on surfaces},\ }\href
  {https://link.aps.org/doi/10.1103/RevModPhys.91.041001} {\bibfield  {journal}
  {\bibinfo  {journal} {Rev. Mod. Phys.}\ }\textbf {\bibinfo {volume} {91}},\
  \bibinfo {pages} {041001} (\bibinfo {year} {2019})}\BibitemShut {NoStop}%
\bibitem [{\citenamefont {Qi}\ \emph {et~al.}(2006)\citenamefont {Qi},
  \citenamefont {Wu},\ and\ \citenamefont {Zhang}}]{qiwuzhang2006prb}%
  \BibitemOpen
  \bibfield  {author} {\bibinfo {author} {\bibfnamefont {X.-L.}\ \bibnamefont
  {Qi}}, \bibinfo {author} {\bibfnamefont {Y.-S.}\ \bibnamefont {Wu}},\ and\
  \bibinfo {author} {\bibfnamefont {S.-C.}\ \bibnamefont {Zhang}},\ }\bibfield
  {title} {\bibinfo {title} {Topological quantization of the spin hall effect
  in two-dimensional paramagnetic semiconductors},\ }\href
  {https://link.aps.org/doi/10.1103/PhysRevB.74.085308} {\bibfield  {journal}
  {\bibinfo  {journal} {Phys. Rev. B}\ }\textbf {\bibinfo {volume} {74}},\
  \bibinfo {pages} {085308} (\bibinfo {year} {2006})}\BibitemShut {NoStop}%
\bibitem [{\citenamefont {Brey}\ \emph {et~al.}(2007)\citenamefont {Brey},
  \citenamefont {Fertig},\ and\ \citenamefont {Das~Sarma}}]{brey2007diluted}%
  \BibitemOpen
  \bibfield  {author} {\bibinfo {author} {\bibfnamefont {L.}~\bibnamefont
  {Brey}}, \bibinfo {author} {\bibfnamefont {H.~A.}\ \bibnamefont {Fertig}},\
  and\ \bibinfo {author} {\bibfnamefont {S.}~\bibnamefont {Das~Sarma}},\
  }\bibfield  {title} {\bibinfo {title} {Diluted graphene antiferromagnet},\
  }\href {https://link.aps.org/doi/10.1103/PhysRevLett.99.116802} {\bibfield
  {journal} {\bibinfo  {journal} {Phys. Rev. Lett.}\ }\textbf {\bibinfo
  {volume} {99}},\ \bibinfo {pages} {116802} (\bibinfo {year}
  {2007})}\BibitemShut {NoStop}%
\bibitem [{\citenamefont {Sherafati}\ and\ \citenamefont
  {Satpathy}(2011)}]{sherafati2011doped}%
  \BibitemOpen
  \bibfield  {author} {\bibinfo {author} {\bibfnamefont {M.}~\bibnamefont
  {Sherafati}}\ and\ \bibinfo {author} {\bibfnamefont {S.}~\bibnamefont
  {Satpathy}},\ }\bibfield  {title} {\bibinfo {title} {Analytical expression
  for the rkky interaction in doped graphene},\ }\href
  {https://link.aps.org/doi/10.1103/PhysRevB.84.125416} {\bibfield  {journal}
  {\bibinfo  {journal} {Phys. Rev. B}\ }\textbf {\bibinfo {volume} {84}},\
  \bibinfo {pages} {125416} (\bibinfo {year} {2011})}\BibitemShut {NoStop}%
\bibitem [{\citenamefont {Bouzerar}(2021)}]{bouzerar2021lieb}%
  \BibitemOpen
  \bibfield  {author} {\bibinfo {author} {\bibfnamefont {G.}~\bibnamefont
  {Bouzerar}},\ }\bibfield  {title} {\bibinfo {title} {{RKKY} couplings in the
  lieb lattice: Flat-band induced frustration},\ }\href
  {https://link.aps.org/doi/10.1103/PhysRevB.104.155151} {\bibfield  {journal}
  {\bibinfo  {journal} {Phys. Rev. B}\ }\textbf {\bibinfo {volume} {104}},\
  \bibinfo {pages} {155151} (\bibinfo {year} {2021})}\BibitemShut {NoStop}%
\bibitem [{\citenamefont {Luo}\ and\ \citenamefont {Yang}(2024)}]{luo2024fano}%
  \BibitemOpen
  \bibfield  {author} {\bibinfo {author} {\bibfnamefont {Y.-D.}\ \bibnamefont
  {Luo}}\ and\ \bibinfo {author} {\bibfnamefont {M.-F.}\ \bibnamefont {Yang}},\
  }\bibfield  {title} {\bibinfo {title} {Influence of flat bands on {RKKY}
  interaction: Perspective of fano defects},\ }\href
  {https://link.aps.org/doi/10.1103/PhysRevB.110.144402} {\bibfield  {journal}
  {\bibinfo  {journal} {Phys. Rev. B}\ }\textbf {\bibinfo {volume} {110}},\
  \bibinfo {pages} {144402} (\bibinfo {year} {2024})}\BibitemShut {NoStop}%
\end{thebibliography}%

\end{document}